\documentclass[prd,aps,preprint,tightenlines,showpacs,nofootinbib,superscriptaddress]{revtex4-1}
\usepackage{mathrsfs}
\usepackage{amsfonts}
\usepackage{amsmath}
\usepackage{amssymb}
\usepackage{array}
\usepackage{verbatim}
\usepackage{bm}
\usepackage{epsfig}
\usepackage{graphicx,color}
\usepackage{relsize}
\usepackage{lineno}
\usepackage{float}
\usepackage{multirow}
\RequirePackage{xspace}
\begin{document}

\title{Exclusive heavy vector meson photoproduction in $pp$  and $pPb$ collisions within the GPD approach:  A phenomenological analysis}
\author{Ya-Ping Xie}
\email{xieyaping@impcas.ac.cn}
\affiliation{Institute of Modern Physics, Chinese Academy of Sciences, Lanzhou 730000, China}
\affiliation{University of Chinese Academy of Sciences, Beijing 100049, China}
\affiliation{State Key Laboratory of Heavy Ion Science and Technology, Institute of Modern Physics,
	Chinese Academy of Sciences, Lanzhou 730000, China}

\author{Victor P. Gon\c{c}alves}
\email{barros@ufpel.edu.br}
\affiliation{Institute of Physics and Mathematics, Federal University of Pelotas, \\
	Postal Code 354,  96010-900, Pelotas, RS, Brazil}

\begin{abstract}
    The exclusive $J/\psi$ and $\Upsilon$ photoproduction in proton - proton ($pp$) and proton - lead ($pPb$) collisions at the LHC energies is investigated using the Generalized Parton Distribution (GPD) approach.  Assuming the Goloskokov - Kroll (GK) model, we estimate the corresponding total cross - sections and rapidity distributions considering different parametrizations for the unpolarized gluon distribution of the proton. In particular, we compare the linear predictions, associated with DGLAP evolution,  with those derived taking into account of the leading nonlinear corrections resulting from gluon recombination. The dependence of our predictions on the choice for the factorization scale is discussed.
\end{abstract}
\maketitle

\section{Introduction}\label{intro}

The possibility of improving our understanding of the quantum 3D imaging of the partons inside the protons and nuclei, and constrain the behavior of the
 gluon distribution at high energies have strongly motivated the study of the exclusive vector meson photoproduction in ultraperipheral collisions (UPCs) at
  the LHC \cite{upc01,upc02,upc03}. Over the last years, this process has been analyzed considering distinct theoretical approaches, based on different
   assumptions and treatments for the meson wavefunctions and descriptions of the QCD dynamics (See, e.g. Refs.
    \cite{Ryskin:1992ui,klein,gluon,Frankfurt:2001db,Ivanov:2004vd,Goncalves:2005yr,Lappi:2013am,Jones:2013pga,
    	 Jones:2015nna,Jones:2016ldq,Jones:2016icr,run2,Flett:2019pux,Mantysaari:2021ryb,Eskola:2022vpi,Eskola:2023oos,Flett:2024htj,Guzey:2024gff,
    	 Luszczak:2024kgi,Goloskokov:2024egn,Xie:2025sfx}). In particular, in Refs.
     \cite{Eskola:2022vpi,Eskola:2023oos,Flett:2024htj,Goloskokov:2024egn,Xie:2025sfx},   the authors have considered the collinear factorization including
      the scale evolution of the Generalised Parton Distributions (GPDs), which we will denote GPD approach hereafter. Such studies have demonstrated the
       importance of the exclusive heavy meson photoproduction to advance our understanding of the 3D content of the proton.

\begin{figure}[t]
 \begin{tabular}{ccc}
 	\includegraphics[width=8cm]{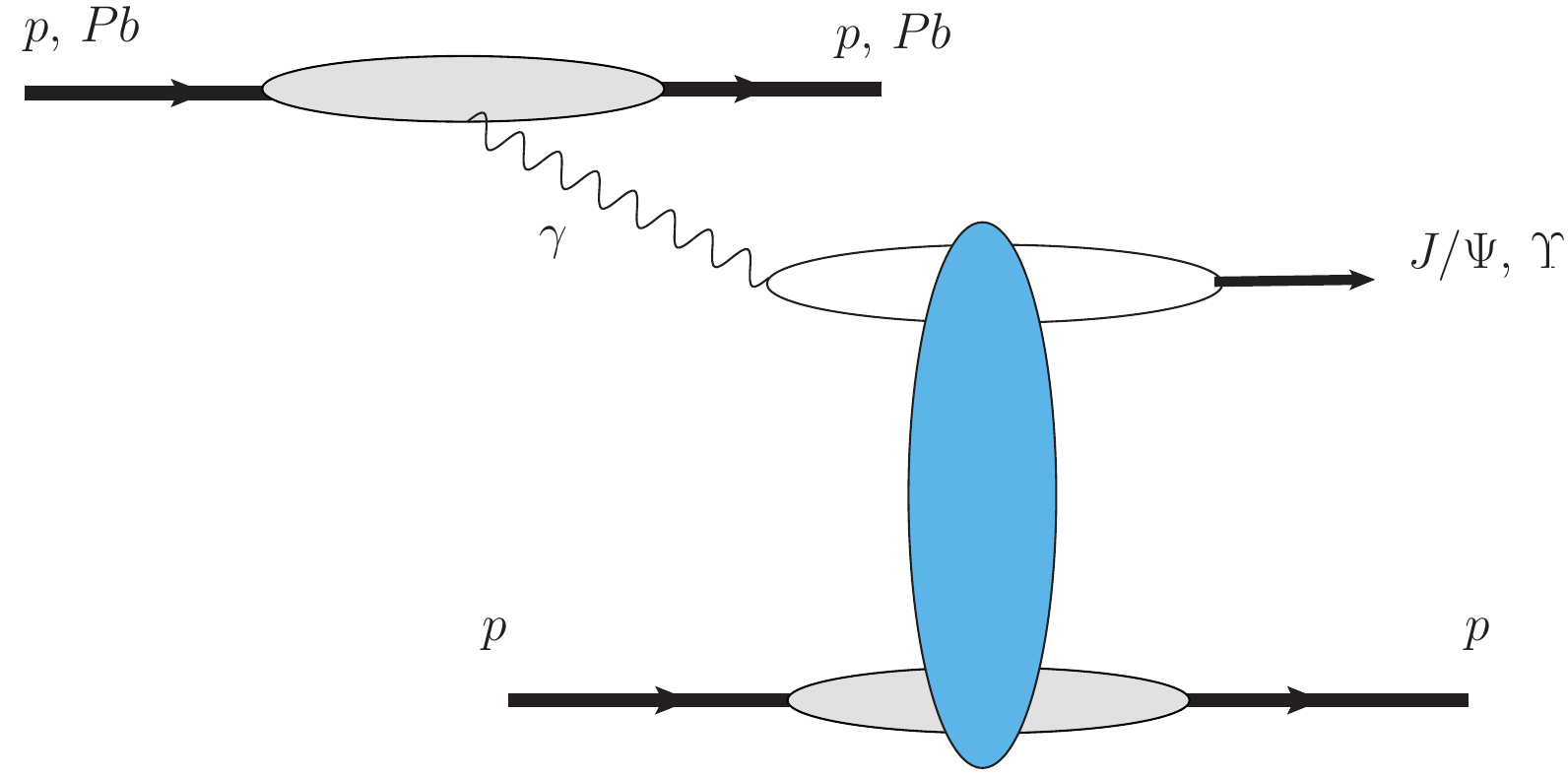} & \,&
	\includegraphics[width=6cm]{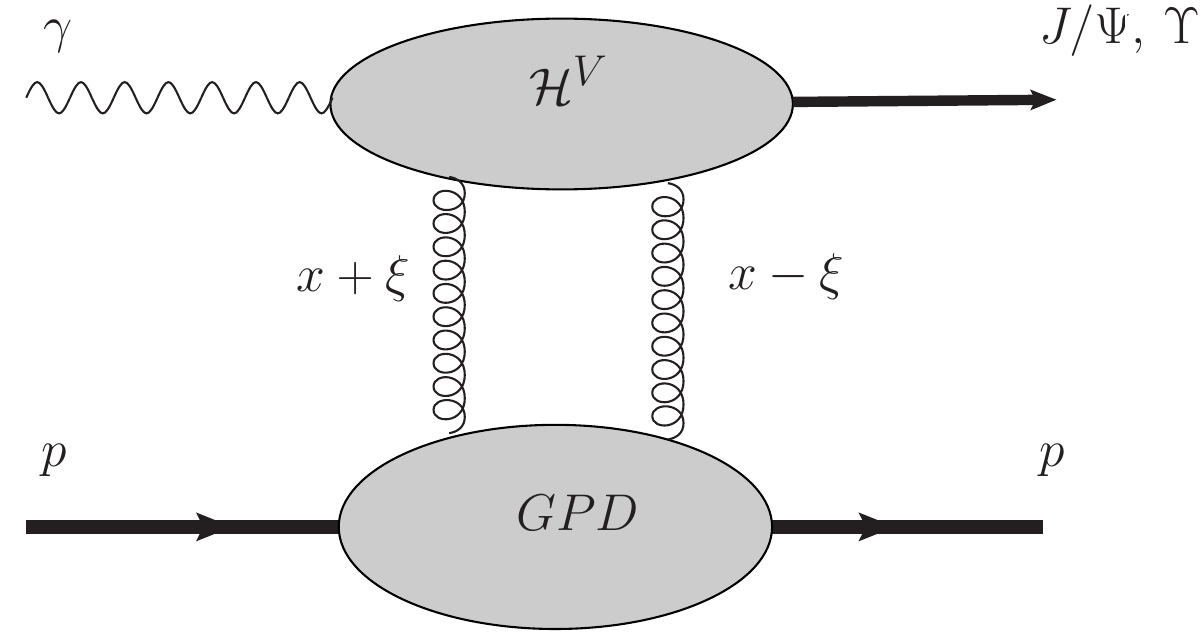} \\
	(a) &\,& (b)
	
 \end{tabular}
	\noindent\caption{(a) Exclusive heavy vector meson photoproduction in $pp$ and $pPb$ collisions. (b) Scattering amplitude for the $\gamma p$ interaction within the GPD approach.}
	\label{Fig:diagram}
\end{figure}

In this paper, we will assume the GPD approach and investigate the exclusive $J/\psi$ and $\Upsilon$ photoproduction in proton - proton ($pp$) and proton
 - lead ($pPb$) collisions at the LHC energies. The process is represented in Fig. \ref{Fig:diagram} (a).  We have that the cross - section for the
  ultraperipheral hadronic collision can be described in terms of the photon flux associated with the incident proton or lead nucleus and the photon - proton
   cross - section. On the hand, the  amplitude for the $\gamma p$ interaction can be factorized in terms of the a hard part that describes the transition of the
  photon into a vector meson and the GPDs that depend on longitudinal momentum of proton ($x$), the skeweness ($\xi$) and the momentum transfer $t$
   [See Fig. \ref{Fig:diagram} (b)]. In our study, we will assume the Goloskokov - Kroll (GK) model
    \cite{Goloskokov:2005sd,Goloskokov:2006hr,Kroll:2012sm}, that allow us to express the gluon GPD in terms of the standard gluon PDF, and derive
     the corresponding total cross - sections and rapidity distributions assuming different parametrizations for the unpolarized gluon distribution of the proton.  
   Our analysis is motivated by the results derived in Refs. \cite{Goloskokov:2024egn,Xie:2025sfx}, which have demonstrated that the data for the $J/\psi$
   production at HERA and $pp$ collisions at the LHC can be described using this approach. In particular, we will extend these previous studies for $pPb$ 
  collisions and for the $\Upsilon$ production. Moreover, we will present, for the first time, the predictions associated with a gluon distribution derived taking 
  into account of the leading nonlinear corrections to the DGLAP evolution resulting from gluon recombination obtained in Ref. \cite{Duwentaster:2023mbk}. 
  Finally,  as our calculation will be performed at leading order, we also will investigate the dependence of our predictions on the choice for the factorization scale assumed in the calculation of the cross - 
  sections.

 This manuscript is organized as follows. In next Section,  we will present a brief review of the formalism needed to describe the exclusive vector meson 
 photoproduction in ultraperipheral hadronic collisions within the GPD approach. In addition, we will discuss the main ingredients used in our calculations. In 
 Section \ref{sec:res} we will present our predictions for the $\gamma p$ cross - sections and for the rapidity distributions associated with the exclusive 
 $J/\psi$ and $\Upsilon$ photoproduction in $pp$ and $pPb$ collisions at the LHC energies, derived assuming different parametrizations for the unpolarized 
 gluon distribution of the proton. A comparison with the current experimental data will be performed. Moreover, we also will estimate the impact of different 
 choices for the factorization scale on  the results  derived using the gluon distribution obtained in Ref. \cite{Duwentaster:2023mbk}, which takes into 
 account of the  leading nonlinear corrections to the QCD dynamics.
Finally, in Section \ref{sec:sum}, we will summarize our main results and conclusions.

\section{Formalism}
In the equivalent photon approximation (EPA) \cite{epa}, the rapidity distribution for the exclusive heavy vector meson photoproduction in proton - hadron ($ph$) interactions, with $h = p$ or $Pb$, can be factorized in terms of the photon flux of one of the incident hadrons  and the photoproduction cross - section as follows 
 \begin{eqnarray}
\frac{d\sigma ^{p+h\to p+h+V}}{d\mathrm{y}}= S_{ph}^2(W_{+})\Big(k_+\frac{dn_p}{dk_+}\Big)\sigma_+(\gamma h \rightarrow V h) +S^2_{ph}(W_{-})\Big(k_-\frac{dn_h}{dk_-}\Big)\sigma_-(\gamma p \rightarrow V p)\,\,,
\label{Eq:rapdis}
\end{eqnarray}
where  we have considered the possibility that both incident hadrons can be source of photons. The factor $S^2_{ph}$ represents the survival factor, which 
is energy dependent, and takes into account for  the probability that the rapidity gap between the vector meson and the hadron target is not populated by 
soft interactions that would destroy the exclusivity of the event.  In our analysis of $pp$ collisions, we will use the survival factors derived in Refs. 
\cite{Jones:2016icr,Flett:2021fvo}. In contrast, for $pPb$ collisions, we will assume  $S_{pPb}^2 = 1$, which is expected to be a reasonable first approximation.   Moreover,  the photon energies $k_i$ can be expressed in terms of the rapidity $y$ and mass $M_V$ of the vector 
meson by the relation $k_{\pm} = (M_V/2)e^{\pm \mathrm{y}}$. Moreover, the photon - hadron center - of - mass energies are given by 
 $W^2_{\pm}=M_V\sqrt{s_{NN}}e^{\pm \mathrm{y}}$. 
 The flux of photons emitted from a proton is given by \cite{Drees:1988pp}
\begin{eqnarray}
\frac{dn_p}{dk}(k)=\frac{\alpha_{em}}{2\pi k}\Big[1+\Big(1-\frac{2k}{\sqrt{s}}\Big)^2\Big]\Big(
\ln \Omega-\frac{11}{6}+\frac{3}{\Omega}-\frac{3}{2\Omega^2}+\frac{1}{3\Omega^3}\Big)\,\,,
\end{eqnarray}	
with $\Omega = 1 + 0.71 \mbox{GeV$^2$}/Q^2_{min}$, where  $Q^2_{min} \approx (k/\gamma_L)^2$ and $\gamma_L$ is the Lorentz factor. 
On the other hand, for a  nucleus, one has \cite{upc02}
\begin{eqnarray}
\frac{dn_{A}}{dk}(k)=\frac{2 Z^2\alpha_{em}}{\pi k}\big[\xi K_1(\xi)
K_0(\xi)-\frac{\xi^2}{2}[K_1^2(\xi)-K_0^2(\xi)]\big],
\end{eqnarray}
where $\xi = k b_{min} / \gamma_L$ with $b_{min}= R_p+R_A$ and $K_0(x)$ and 
$K_1(x)$ are modified Bessel functions.
 As the photoproduction cross - section increases with the energy, the first (second) term in Eq. (\ref{Eq:rapdis}) is the dominant contribution for positive 
 (negative) rapidities, when we have the collision between identical particles. For asymmetric collisions, as e.g. in $pPb$ collisions, the dependence of the 
 photon flux on the nuclear charge implies that the contribution associated with $\gamma p$ interactions become dominant.

The total cross - section for the exclusive vector meson production in photon - hadron interactions is given by
\begin{eqnarray}
\sigma^{\gamma+p\to V+p}(W)=\int d t \frac{d\sigma_T}{dt} =\int d t \frac{1}{16\pi W^4}|{\cal{A}}(W,t)|^2 \,\,,
\end{eqnarray}
where the scattering amplitude $\mathcal{A}$ is represented in Fig. \ref{Fig:diagram} (b). We have that the two exchanged gluons  carry different fractions 
of the incoming hadron momentum, which implies that the description of the process involves a generalized (skewed) gluon distribution.  As the  formation 
time, $\tau_f \approx 2k/M_V^2$ is much greater than the $q\bar{q}$ - hadron interaction time, the amplitude can be factorized in terms of the function 
$\mathcal{H}^{V}$, which describes the photon - vector meson transition, and the gluon GPDs. In our analysis, we will use the function $\mathcal{H}^{V}$ 
derived in Ref. \cite{Goloskokov:2024egn}  at leading order.  In particular, we have that the squared scattering amplitude can be expressed  as follows
\begin{eqnarray}
|{\cal{A}}(W,t)|^2 = \big[|\mathcal{M}_{++,++}|^2+|\mathcal{M}_{+-,++}|^2\big] \,\,,
\end{eqnarray}
where  the helicity amplitudes ${\cal{M}}$ are defined by~\cite{Goloskokov:2024egn}
\begin{eqnarray}
\mathcal{M}_{\mu^\prime+,\mu+}
&=& \frac{e}{2}C_V\int_0^1\frac{dx}{(x+\xi)(x-\xi+i\epsilon)}\nonumber\\
&&\times\big\{\mathcal{H}^{V+}_{\mu^\prime,\mu}\,H_g(x, \xi, t,\mu_F)   +\mathcal{H}^{V-}_{\mu^\prime,\mu}\,\tilde{H}_g(x, \xi, t, \mu_F]\big\}\,\,,
\end{eqnarray}
and
\begin{eqnarray}
\mathcal{M}_{\mu^\prime-,\mu+}
&=&-\frac{e}{2}C_V\frac{\sqrt{-t}}{2m}\int_0^1\frac{dx}{(x+\xi)(x-\xi+i\epsilon)}\nonumber\\
&&\times\big\{\mathcal{H}^{V+}_{\mu^\prime,\mu}\,E_g(x, \xi, t,\mu_F)
+\mathcal{H}^{V-}_{\mu^\prime,\mu}\,\tilde{E}_g(x, \xi, t,\mu_F)\big\} \,\,,
\label{eq:ampliM}
\end{eqnarray}
with  $H_g(x, \xi, t, \mu_F)$ and $E_g(x, \xi, t, \mu_F)$ [$\tilde{H}_g(x, \xi, t,\mu_F)$ and  $ \tilde{E}_g(x, \xi, t,\mu_F)$]  being the unpolarized [polarized] gluon GPDs.  In the vector meson production, the dominant contribution is $H_g(x, \xi, t, \mu_F)$, thus, the other GPD functions can
 be neglected in the following calculations.
The flavor factor $C_V$ is equal to 2/3 (1/3) for the $J/\psi $ ($\Upsilon$) meson. 
The functions $\mathcal{H}^{V\pm}_{\mu^\prime,\mu}$ are determined in terms of a sum or difference between the  amplitudes for different gluon helicities, which implies
\begin{eqnarray}
\mathcal{H}^{V\pm}_{\mu^\prime,\mu} & = &  \big[\mathcal{H}^V_{\mu^\prime+,\mu+}\pm\mathcal{H}^V_{\mu^\prime-,\mu-}\big] \nonumber \\
& = & 64\pi^2\alpha_s(\mu_R)\int_0^1d\tau
\int \frac{d^2\mathbf{k}_\perp}{16\pi^3}\psi^V(\tau, \mathbf{k}_\perp)\mathcal{F}_{\mu^\prime,\mu}^{\pm}(\tau, x, \xi, \mathbf{k}^2_\perp)\,\,,
\end{eqnarray}
where $\alpha_s(\mu^2)$ is QCD running coupling, given by $\alpha_s(\mu^2) = 4\pi / (\beta_0 \log(\mu^2/\Lambda^2)$, with $\Lambda = 0.22$ GeV. In our calculations, we assume $\mu = m_V$ and $m_q = m_V/2$. Moreover, $\psi^V(\tau, \mathbf{k}_\perp)$ is the wave function of the vector meson and $\mathcal{F}_{\mu^\prime,\mu}^{\pm}$ can be  calculated perturbatively taking into account of the tranverse momentum $\mathbf{k}_\perp$ of the quark (and antiquark), 
and is given by~\cite{Goloskokov:2024egn} 
\begin{eqnarray}
\mathcal{F}_{\mu^\prime,\mu}^{\pm}=\frac{f^{\pm}_{\mu^\prime,\mu}}{(2\mathbf{k}^2_\perp+m_V^2)(4\xi\mathbf{k}_\perp^2
	+m_V^2(\xi-x)+i\epsilon)(4\xi\mathbf{k}_\perp^2+m_V^2(\xi+x))} \,\,,
\end{eqnarray}
with$f_{+,+}^+ = 64m^5_V(x^2 - \xi^2)$ and $f_{+,+}^- = -256m_V^3\mathbf{k}_\perp^2x\xi$.
 Following Refs.~\cite{Goloskokov:2024egn,Xie:2025sfx}, the wave function will be written as 
\begin{equation}
\psi^V(\tau, \mathbf{k}_\perp)=N_V\exp\Big(-a_{V}^2\frac{ \mathbf{k}^2_\perp}{\tau(1-\tau)}\Big)\,\,,
\end{equation}
with $\tau = 1/2$. 
In particular, for the $J/\psi$ case, $N_{J/\psi}=0.025$ GeV$^{-1}$. As shown in Ref.~\cite{Xie:2025sfx}, it implies $a_{\psi}^2 = 0.091 $GeV$^{-2}$. In our analysis of the $\Upsilon$ case, we will assume $a_{\Upsilon}^2 = a_{\psi}^2$, and  the value of $N_{\Upsilon}$ will be fixed using the HERA data. In this work, $N_{\Upsilon} = 0.192$ GeV$^{-1}$.

                                                        

In order to estimate the gluon GPDs, present in Eq. (\ref{eq:ampliM}), we 
will consider the model proposed and detailed in Refs. \cite{Goloskokov:2005sd,Goloskokov:2006hr,Kroll:2012sm}, usually called Goloskokov - Kroll (GK) 
model, which is based on fits of meson electroproduction data. In this model,  the gluon GPDs are expressed as follows
\begin{equation}
F_g(x, \xi, t) = \int_1^1d\beta \int_{-1+|\beta|}^{1-|\beta|}d\alpha \, \delta(\beta + \xi \alpha -x)f_g(\beta, \alpha, t).
\end{equation}
where  the double distribution $f_g(\beta, \alpha, t)$ for gluons, introduced originally in Refs.~\cite{Radyushkin:1996ru,Radyushkin:1997ki}, is  given by 
\begin{eqnarray}
f_g(\beta, \alpha, t) = e^{B_V t}  \beta g(\beta,\mu^2) \frac{15}{16} \frac{[(1-|\beta|)^2-\alpha^2]^{2}}{(1-|\beta|)^{5}}.
\label{eq:gluon}
\end{eqnarray}
Here,  $g(\beta,\mu^2)$ is the usual unpolarized gluon PDF at the hard scale $\mu^2$. As a consequence, using this approach, the energy dependence of 
the photon - hadron cross section will be sensitive to the $x$ - dependence of $g(x,\mu^2)$, which is determined by the QCD dynamics. 

In our calculations, we will assume the approximation ${d\sigma}/{dt}\simeq ({1}/{2B_V})
{d\sigma}/{dt}(t=0)$, with the slope $B_V$ determined by  fitting the experimental data. Following Ref. \cite{Goloskokov:2024egn}, we will take for the 
$J/\psi$ case that $
B_V= b_0+\alpha' \ln[W^2/m_V^2]$, with  $\alpha'=0.185$ GeV$^{-2}$ and $b_0=1.06$ {GeV}$^{-2}$. 
For the $\Upsilon$ production, we will we assume $2B_V= 4.63 \mbox{GeV}^{-2} + 0.24\mbox{GeV}^{-2} \log(W/90.0\mbox{GeV})$, as in 
Ref.\cite{Jones:2013pga}. In addition, as discussed above, the normalization of the $\Upsilon$ wave function was adjusted in order to describe the current HERA data.

 \begin{figure}[t]
	\includegraphics[width=8cm]{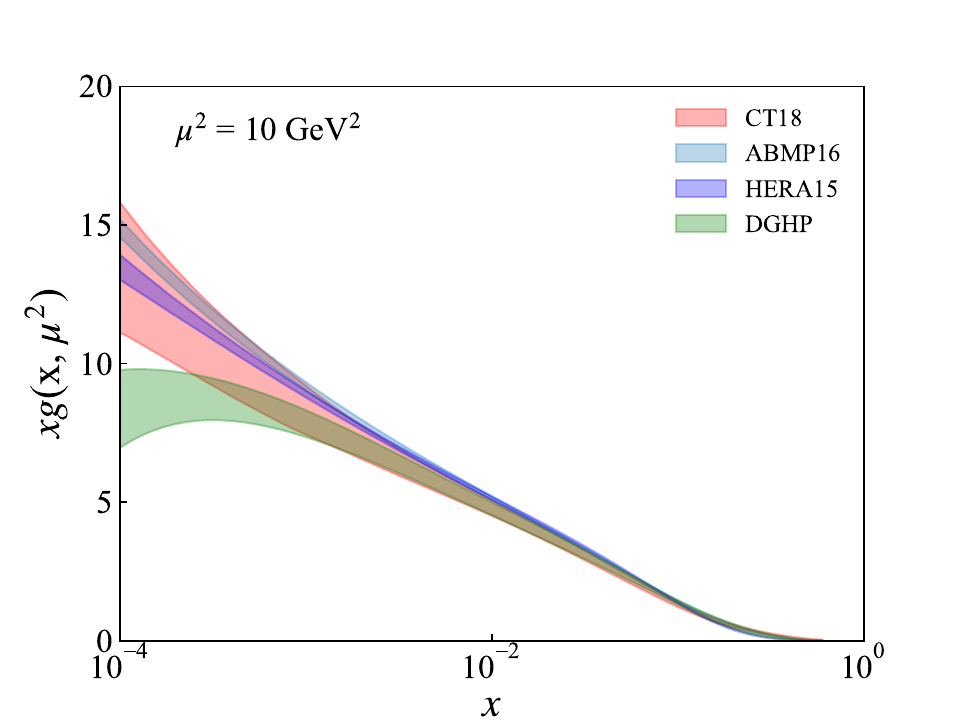}
	\includegraphics[width=8cm]{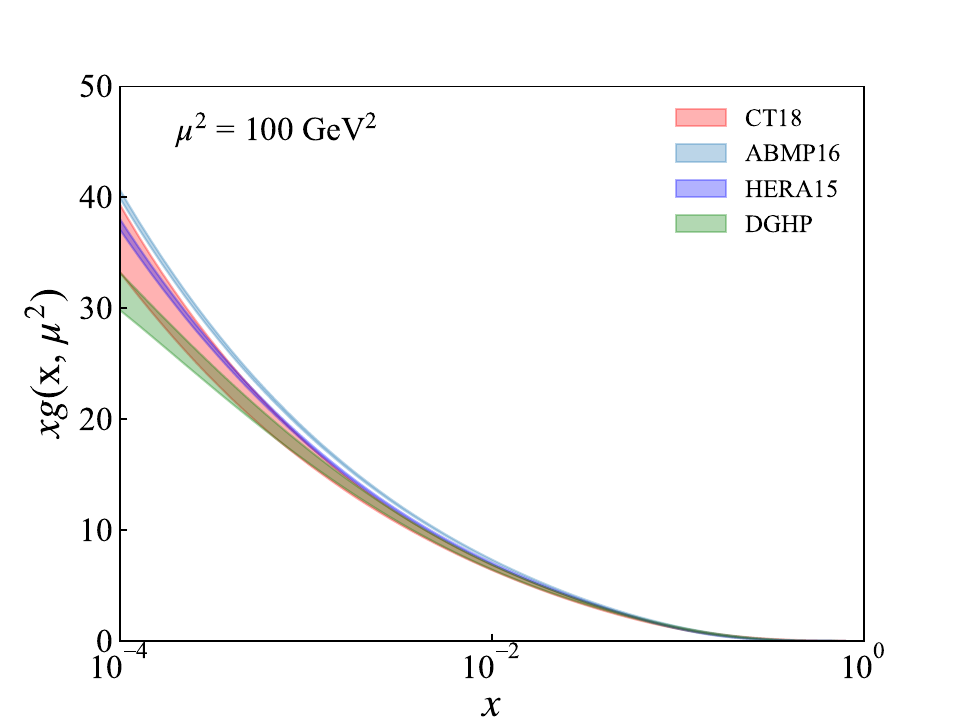}
	\noindent\caption{Predictions of the different PDF sets considered in our study for the $x$ - dependence of the gluon distribution, $xg(x,\mu^2)$, for two 
	different values of the hard scale:  $\mu^2 = 10 $ GeV$^2$ (left panel) and $\mu^2 = 100 $ GeV$^2$ (right panel).}
	\label{fig:gluon}
\end{figure}

The last ingredient in our calculations is the description of the proton gluon distribution, present in Eq. (\ref{eq:gluon}). As in Refs. \cite{Goloskokov:2024egn,Xie:2025sfx}, we will consider 
three different PDF sets,  ABMP16~\cite{Alekhin:2018pai}, CT18~\cite{Hou:2019qau} and HERA15~\cite{H1:2015ubc}, which perform a global fit of the 
current experimental data using  the DGLAP evolution equations. In addition, we will consider the parametrization obtained in 
Ref. \cite{Duwentaster:2023mbk}, denoted DGHP hereafter, which have performed a fitting of the current lepton - proton deep inelastic scattering data using 
nonlinear evolution equations, which take into account of the leading nonlinear corrections to the DGLAP equations associated with the gluon recombination 
effects. In particular, we will use the R1 set for the DGHP parameterization. In Fig. \ref{fig:gluon} we present a comparison between the predictions of these different PDF sets for the gluon distribution, considering two 
different values for the hard scale. We have that the predictions are similar for large $x$, but the inclusion of the nonlinear effects in the QCD dynamics 
implies a smaller DGHP gluon distribution at smaller values of $x$. As expected, the impact of the nonlinear effects decreases with the increasing of the 
hard scale.
 
 \begin{figure}[t]
	\includegraphics[width=8.0cm]{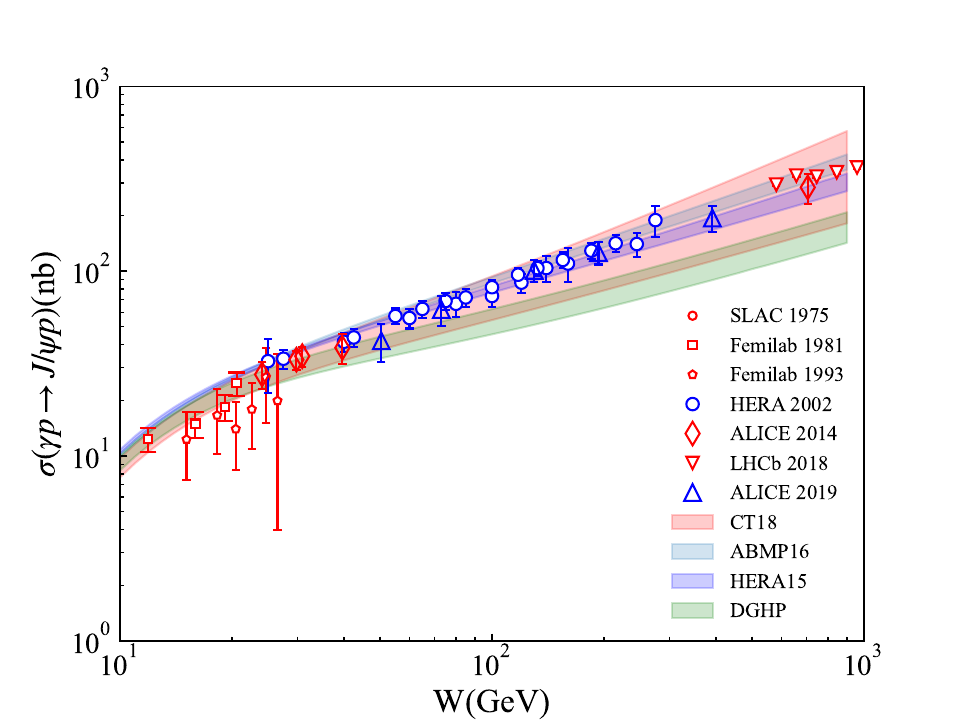}
	\includegraphics[width=8.0cm]{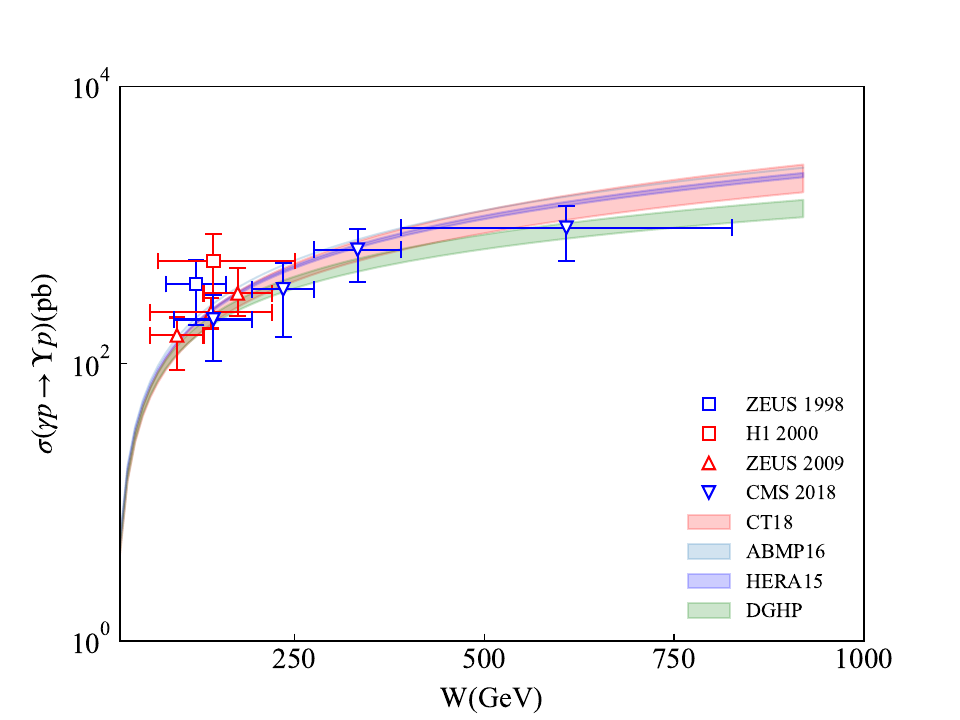}
	\caption{Energy dependence of the exclusive $J/\psi$ (left panel) and $\Upsilon$ (right panel) production cross - section in photon-proton interactions.
	The measured cross sections for the  $J/\psi$  and $\Upsilon$ photoproduction are taken from 
		Refs.\cite{Camerini:1975cy, Binkley:1981kv, Frabetti:1993ux, Adloff:2000vm, Chekanov:2002xi,LHCb:2018rcm,ALICE:2014eof, ALICE:2018oyo} and \cite{ZEUS:1998cdr,H1:2000kis,Chekanov:2009zz,CMS:2018bbk}, respectively.}
		\label{fig:photonproton}
\end{figure}

 \section{Numerical results and discussion}
 \label{sec:res}
 
 In what follows, we will present our predictions for the exclusive $J/\psi$ and $\Upsilon$ photoproduction considering the gluon PDFs discussed in the 
 previous section, considering the associated uncertainties in the parametrizations provided by the different groups. Initially, we will estimate the cross - 
 sections assuming that the hard scale is given the mass of the vector mass ($\mu^2 = m_V^2$). In Fig. \ref{fig:photonproton} we compare our predictions 
 for the energy dependence of the exclusive $J/\psi$ (left panel) and $\Upsilon$ (right panel) production cross - section in photon-proton interactions with the 
 current experimental data~\cite{Camerini:1975cy, Binkley:1981kv, Frabetti:1993ux, Adloff:2000vm, Chekanov:2002xi}. As already observed in 
 Refs.~\cite{Goloskokov:2024egn,Xie:2025sfx}, the predictions associated with the ABMP16~\cite{Alekhin:2018pai}, CT18~\cite{Hou:2019qau} and 
 HERA15~\cite{H1:2015ubc} parametrizations, based on the linear DGLAP equations, are able to describe the $J/\psi$ data. Our results indicate that such a 
 conclusion is also valid in the $\Upsilon$ case. GK model with three linear evolution gluon density give a good agreement to photoproduction of $\Upsilon$
 data \cite{ZEUS:1998cdr,H1:2000kis,Chekanov:2009zz,CMS:2018bbk}.
  On the other hand, the DGHP predictions describe the $\Upsilon$ data and underestimate the $J/\psi$ data 
 at high energies.

\begin{figure}[t]
\includegraphics[width=8cm]{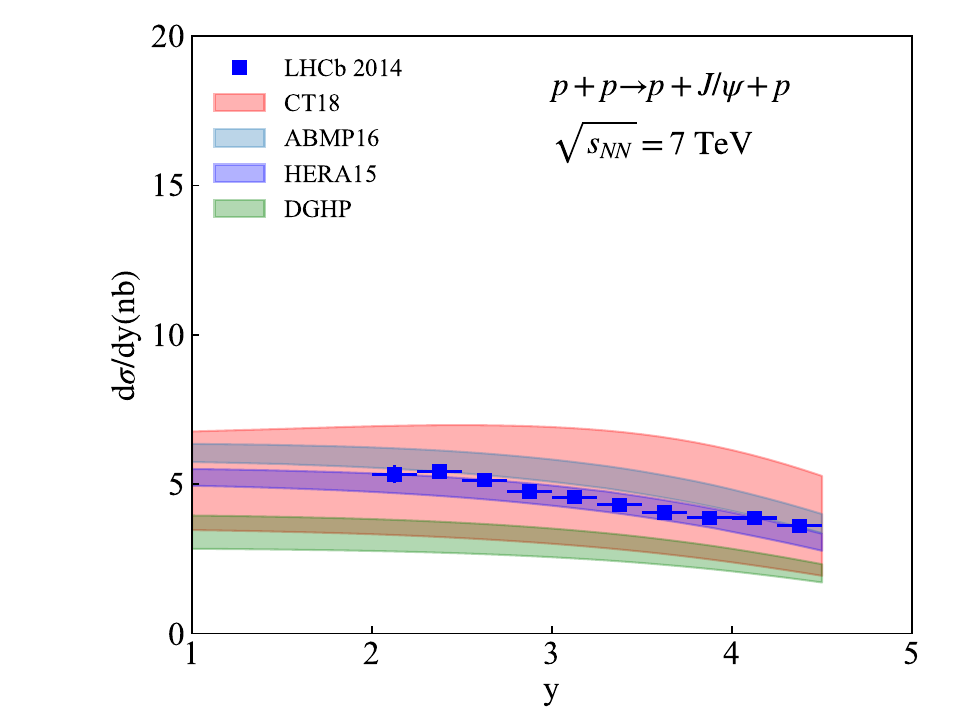}
\includegraphics[width=8cm]{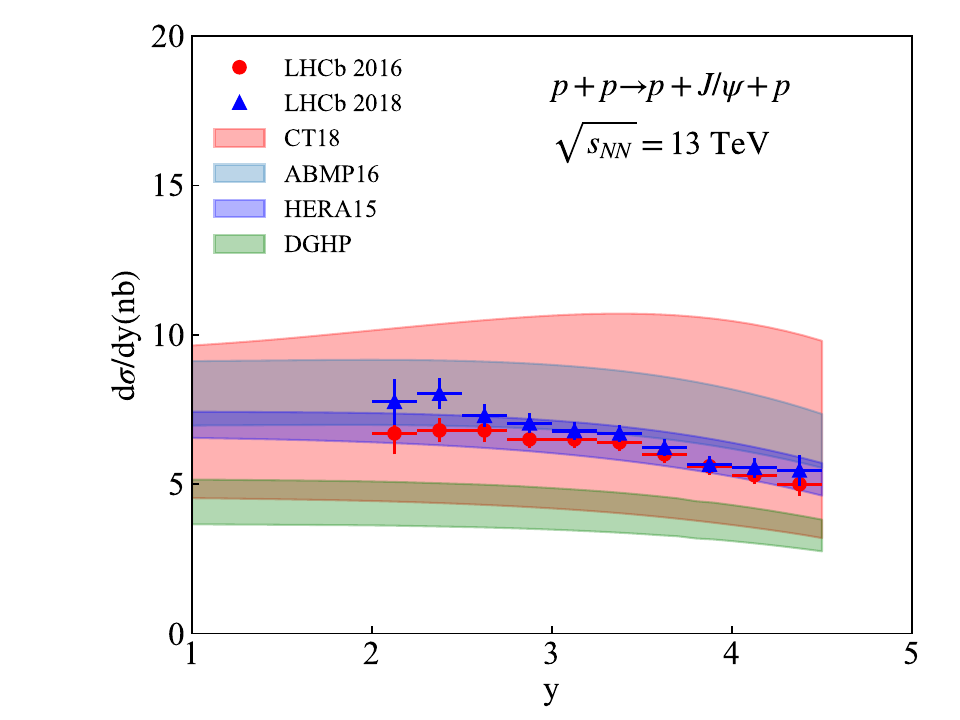}
\includegraphics[width=8cm]{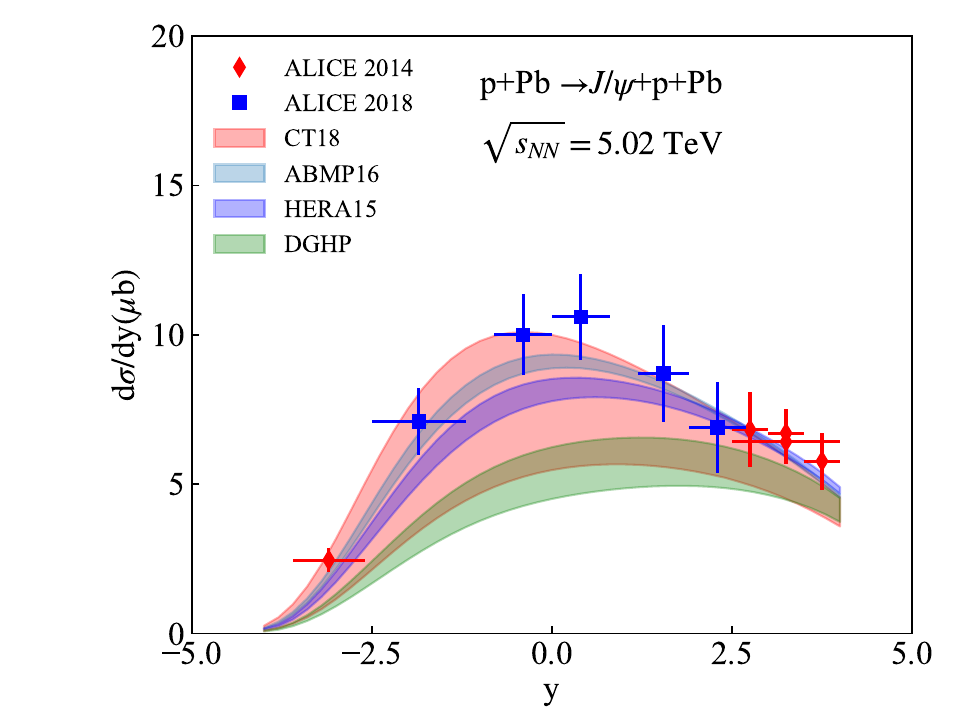}
\includegraphics[width=8cm]{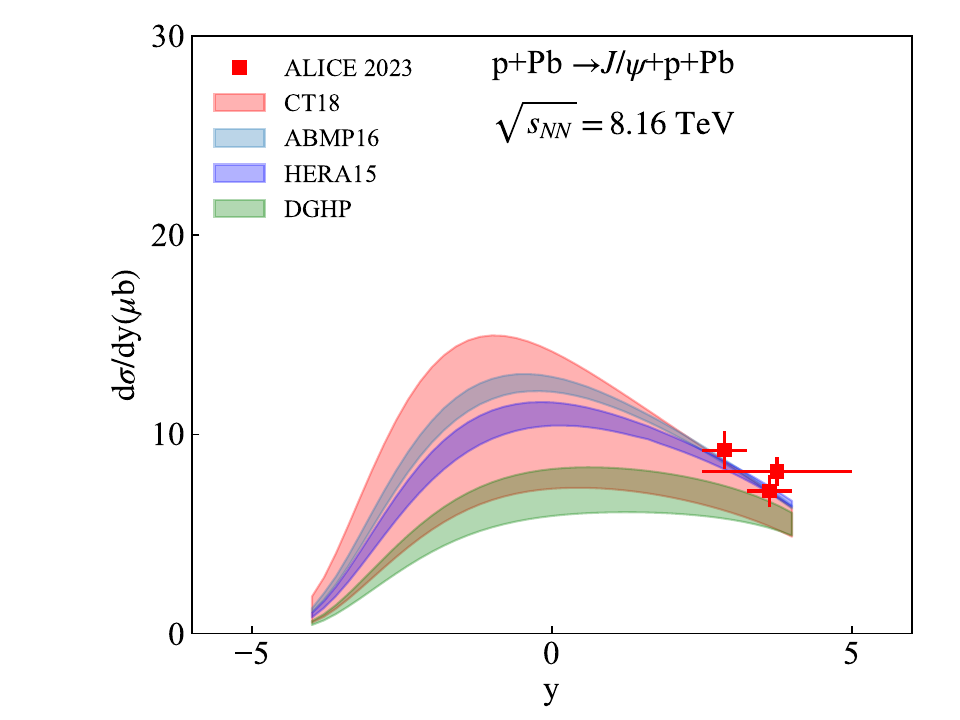}
\noindent\caption{Rapidity distributions for the exclusive $J/\psi$ photoproduction in $pp$ (upper panels) and $pPb$ (lower panels)  collisions. 
Experimental data of pp are taken from 
Refs.\cite{LHCb:2014acg,LHCb:2016oce,LHCb:2018rcm} while pPb are from \cite{ALICE:2014eof,ALICE:2018oyo,ALICE:2023mfc}. }
	\label{fig:rapdist_jpsi}
\end{figure}

\begin{figure}[t]
	\includegraphics[width=8cm]{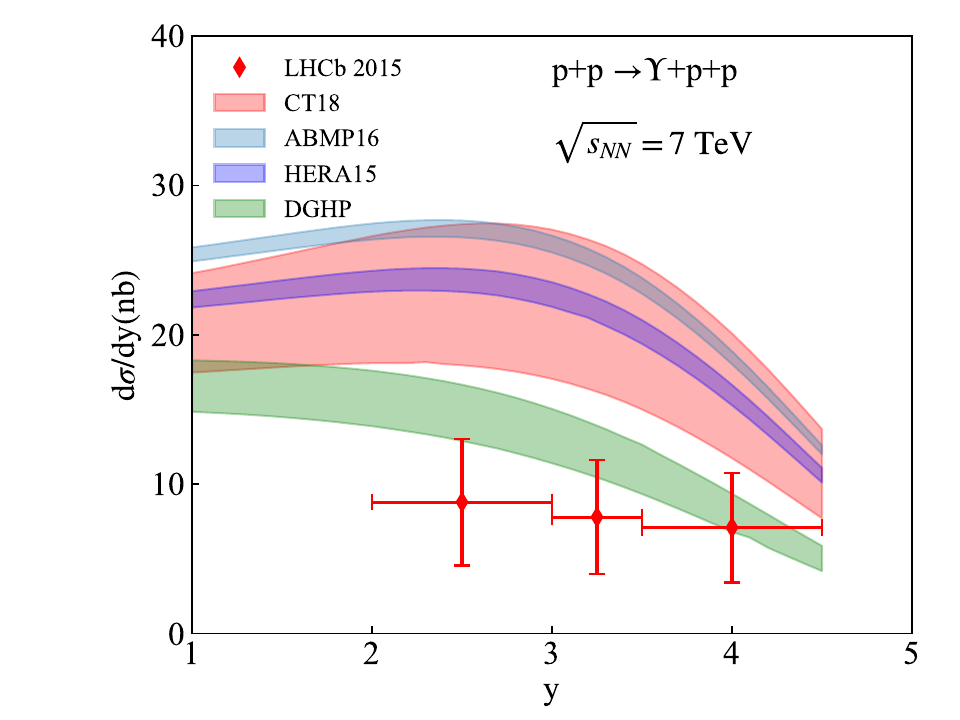}
	\includegraphics[width=8cm]{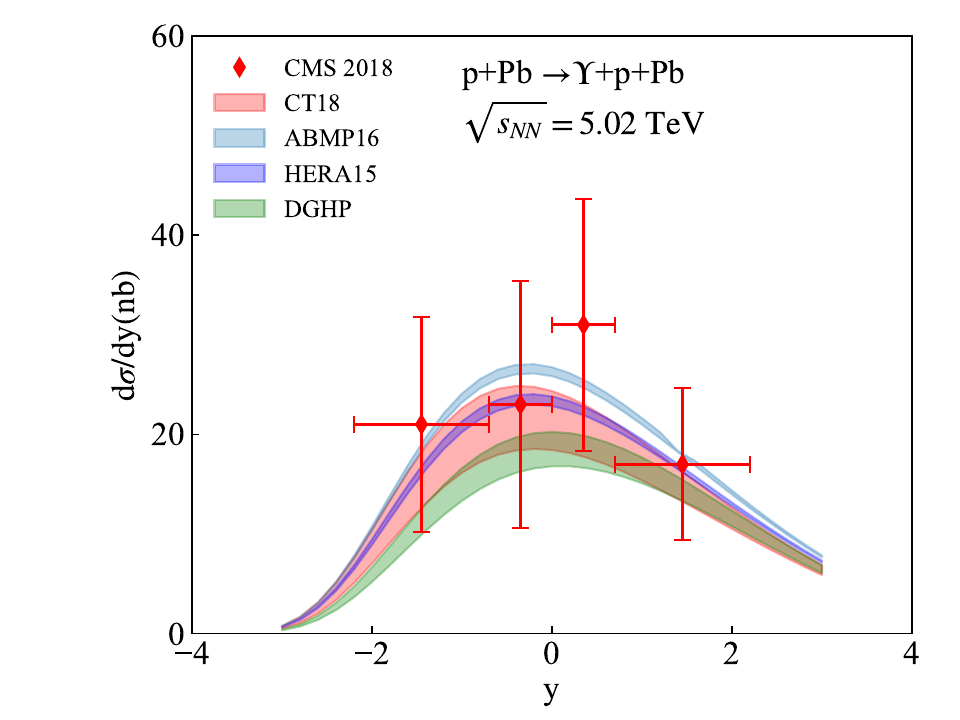}
\noindent\caption{Rapidity distributions for the exclusive $\Upsilon$ photoproduction in $pp$ (left panel) and $pPb$ (right panel) collisions. Experimental data  are taken from Refs \cite{LHCb:2015wlx,CMS:2018bbk}. }
	\label{fig:rapdist_ups}
\end{figure}

In Fig. \ref{fig:rapdist_jpsi} we present a comparison between our predictions for the rapidity distribution of exclusive $J/\psi$ photoproduction  in $pp$ and $pPb$ collisions at the LHC energies with the LHCb \cite{LHCb:2014acg,LHCb:2016oce} and ALICE data
 \cite{ALICE:2014eof,ALICE:2018oyo,ALICE:2023mfc}. In this case,  the contribution associated with $\gamma Pb$ interactions was
  disregarded, since it is very small in the rapidity range of the data. As expected from the results presented in Fig. \ref{fig:photonproton}, the DGLAP
  predictions provide a satisfactory description of the current data, while they are  underestimated when the nonlinear effects, as described by the DGHP
  parametrization, are taken into account. In contrast, the results for the $\Upsilon$ production presented in Fig. \ref{fig:rapdist_ups}, indicate that the linear
  predictions are not able to describe the $pp$ data\cite{LHCb:2015wlx}. For $pPb$ collisions, the current data have a large uncertainty\cite{CMS:2018bbk}, 
  which does not allow discriminating
 between the distinct predictions. As a consequence, a stronger conclusion about the preferred description of the gluon distribution is still not possible.

\begin{figure}[t]
	\includegraphics[width=8cm]{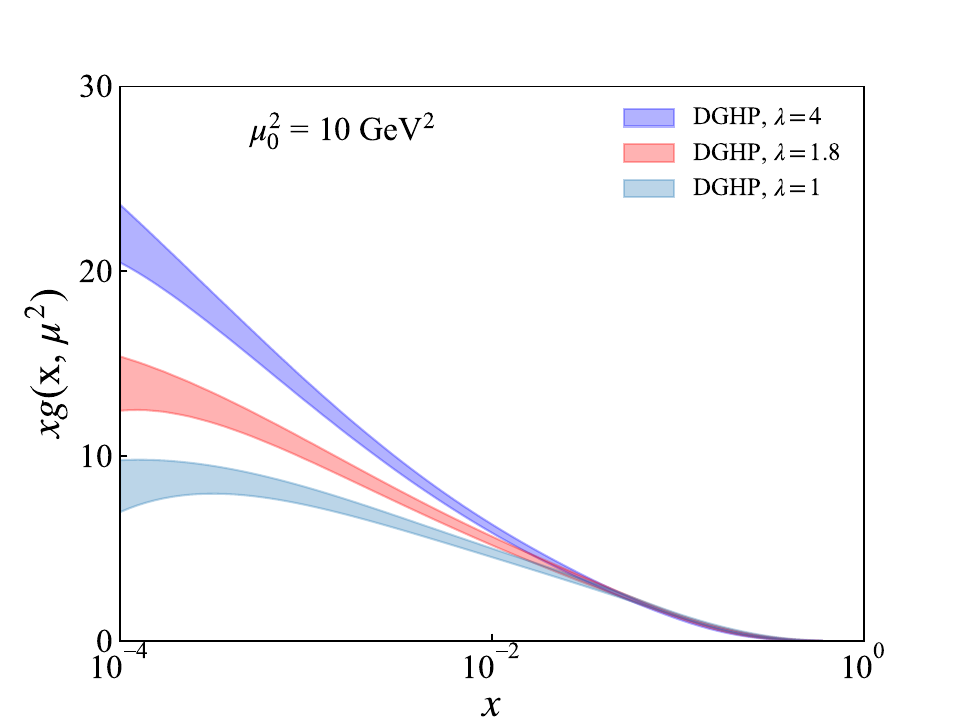}
	\includegraphics[width=8cm]{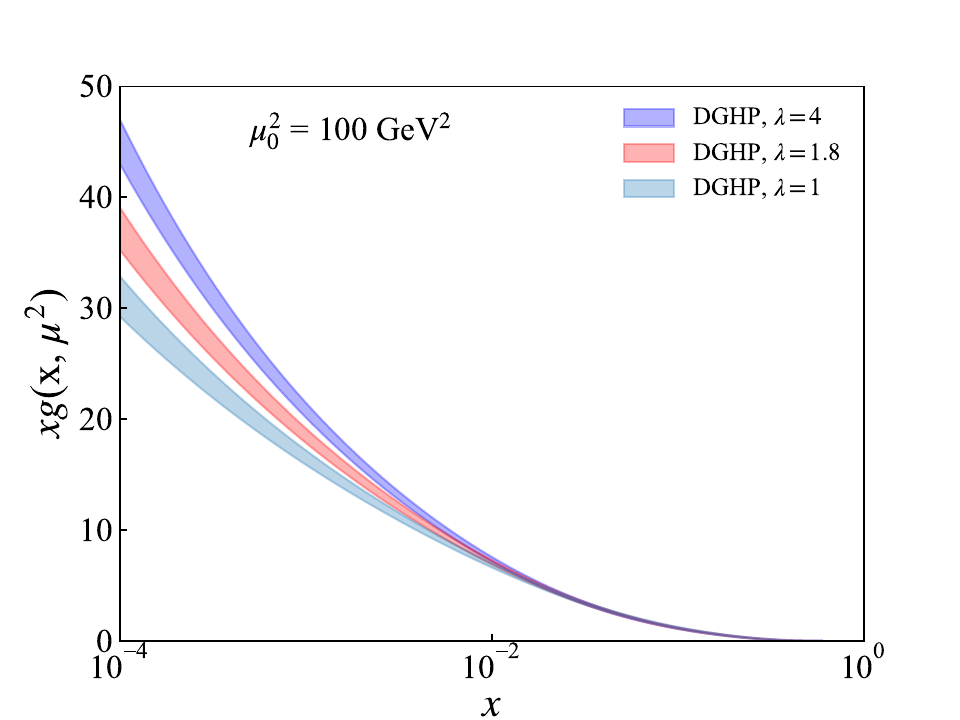}
	\noindent\caption{Predictions of the DGHP parametrization for the gluon distribution evaluated for different values of the hard scale $\mu^2 = \lambda 
		\mu_0^2$. }
		\label{fig:gluon_dghp}
\end{figure}

Previous studies of the exclusive vector meson photoproduction using the GPD model have demonstrated that the associated predictions are strongly 
sensitive to the choice of the hard scale assumed to describe the process\cite{Goloskokov:2024egn}. In order to estimate the dependence of our 
predictions on this choice, in what 
follows we will perform a comparison of the DGHP results, derived assuming $\mu^2 = m_V^2$, with those obtained for $\mu^2 = \lambda m_V^2$, with 
$\lambda$ = 1.8 and 4. 
The impact of these difference values of $\mu^2$ on the gluon distribution are presented in Fig. \ref{fig:gluon_dghp}. We have found that the behavior of the 
gluon distribution at small - $x$ is strongly dependent on the value assumed for $\lambda$, which becomes steeper with the increasing of the hard scale. 
Such a behavior occurs since the contribution of the nonlinear corrections are larger for smaller values of the hard scale. The increasing of $\lambda$ offsets
the nonlinear effect of gluon PDF. 

\begin{figure}[t]
	\includegraphics[width=8cm]{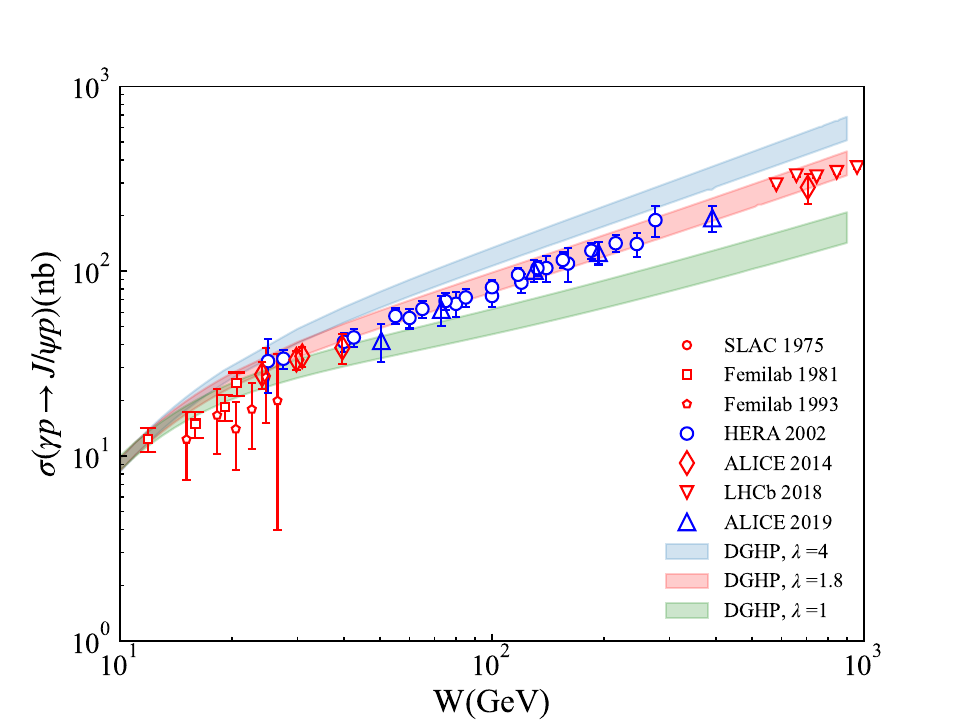}
	\includegraphics[width=8cm]{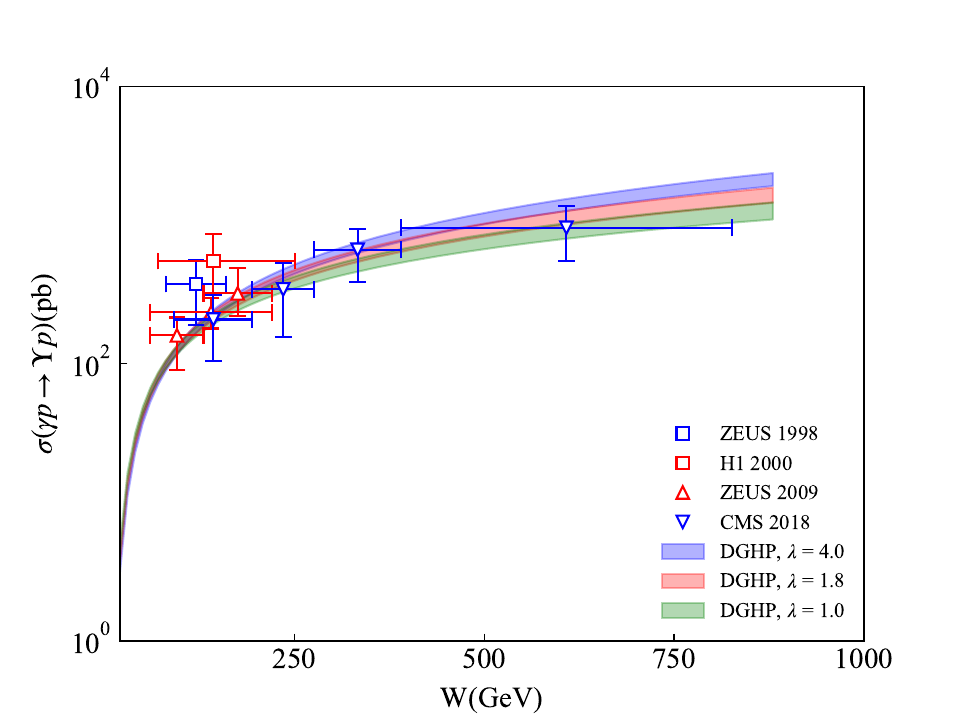}
	\caption{Energy dependence of the exclusive $J/\psi$ (left panel) and $\Upsilon$ (right panel) production cross - section in photon-proton interactions, derived using the DGHP parametrization and different values of the hard scale $\mu^2 = \lambda m_V^2$. 
		The measured cross section of $J/\psi$ are taken from 
		Refs.\cite{Camerini:1975cy, Binkley:1981kv, Frabetti:1993ux, Adloff:2000vm, Chekanov:2002xi,LHCb:2018rcm,ALICE:2014eof, ALICE:2018oyo} and $\Upsilon$ are taken from
		Refs.\cite{ZEUS:1998cdr,H1:2000kis,Chekanov:2009zz,CMS:2018bbk}. }
\label{fig:photonproton_dghp}
\end{figure}

\begin{figure}[t]
\includegraphics[width=8cm]{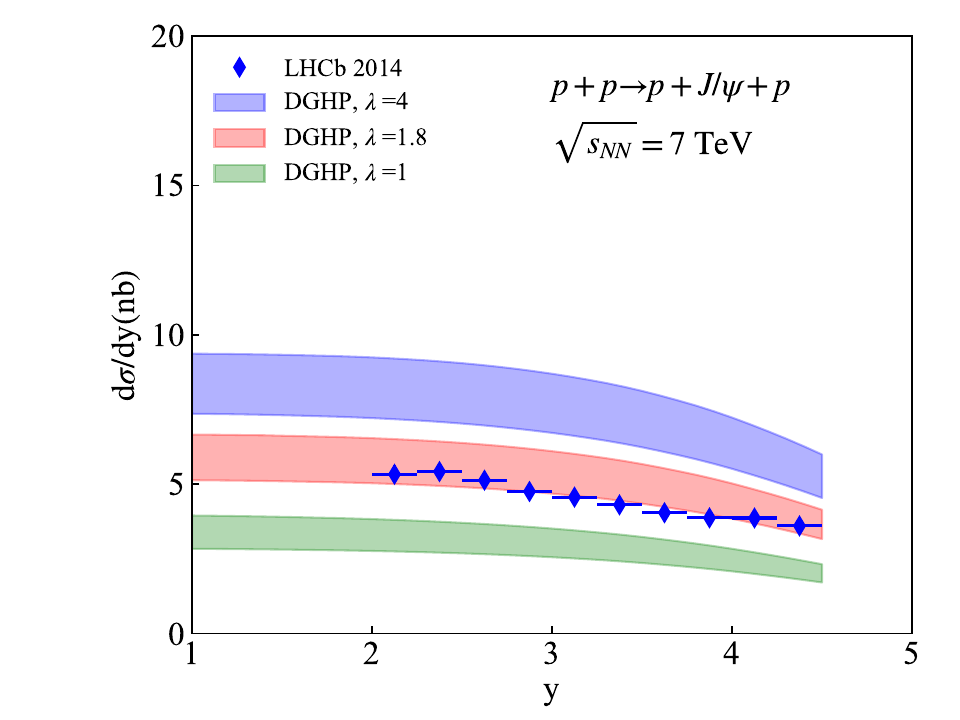}
\includegraphics[width=8cm]{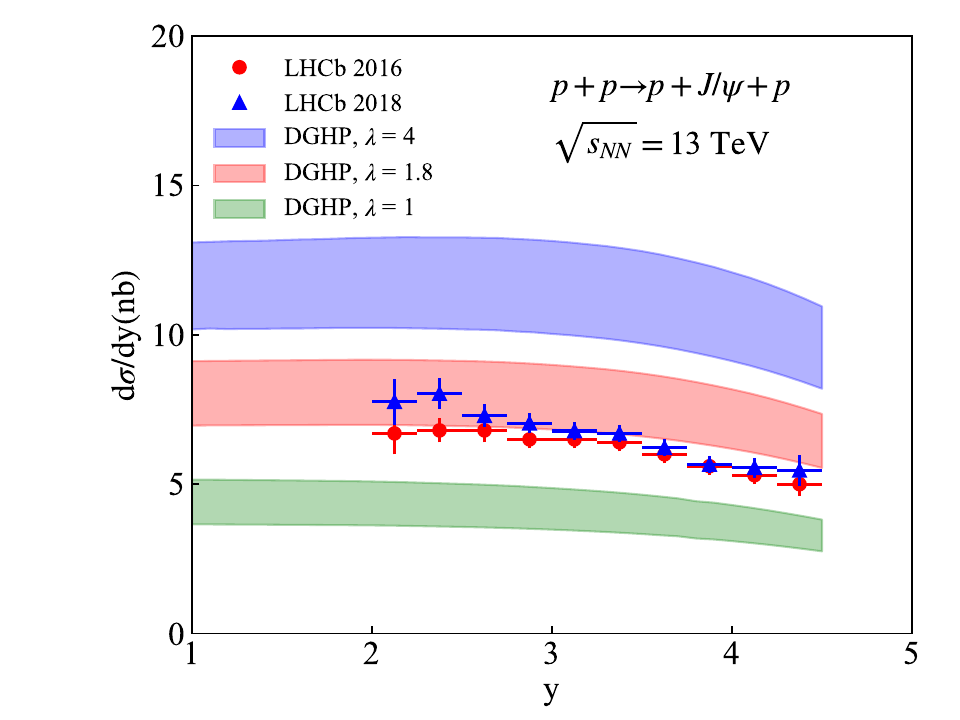}
	\includegraphics[width=8cm]{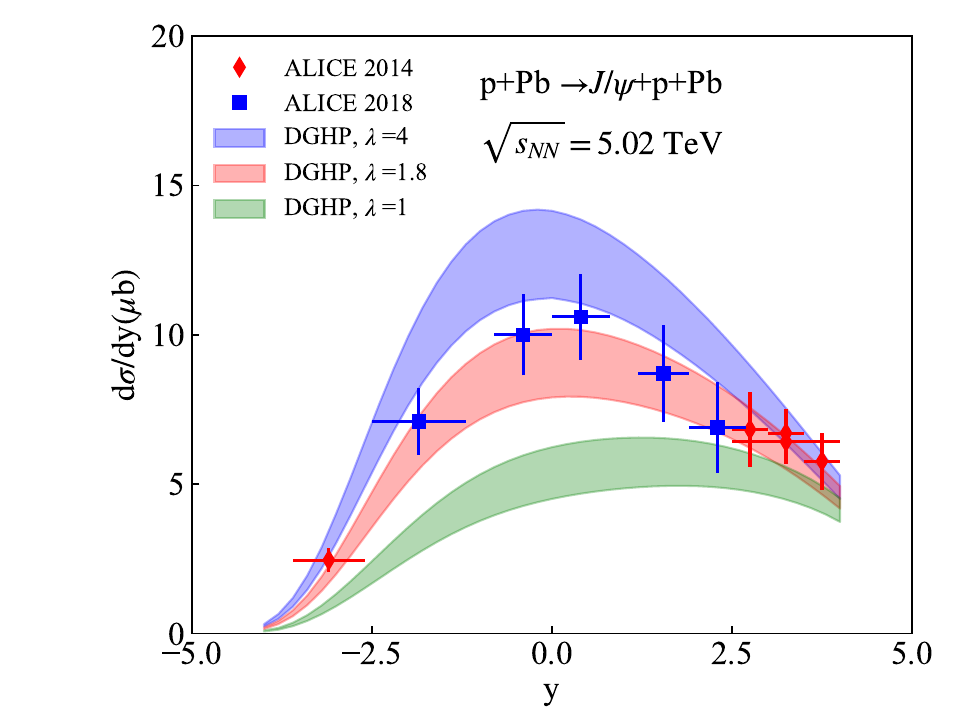}
	\includegraphics[width=8cm]{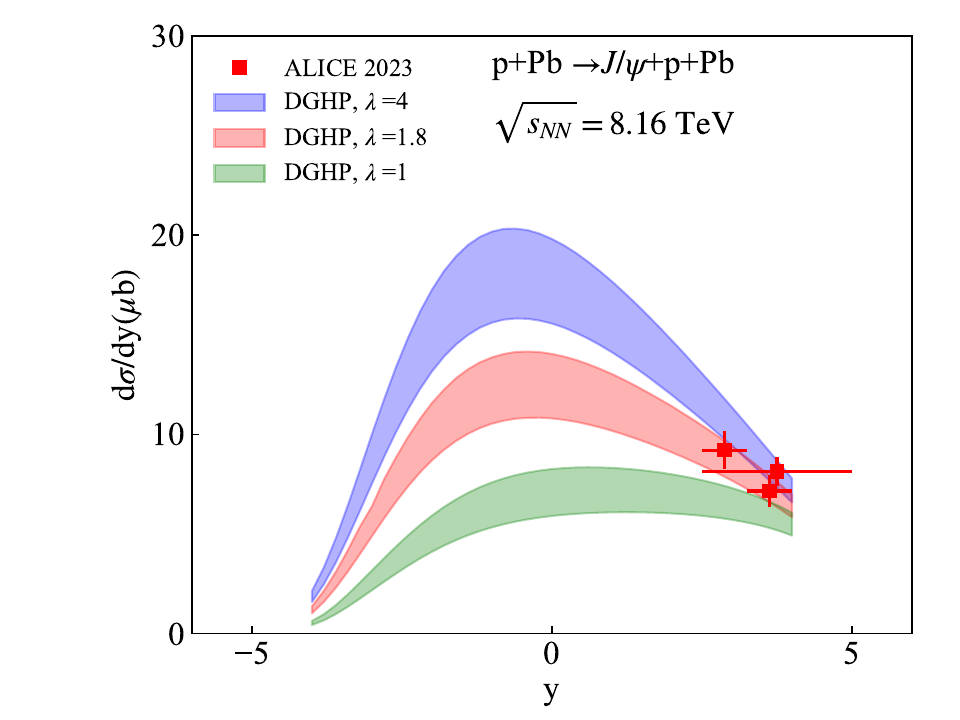}
\noindent\caption{Rapidity distributions for the exclusive $J/\psi$ photoproduction in $pp$ (upper panels) and $pPb$ (lower panels)  collisions,
	 derived using the DGHP parametrization and different values of the hard scale $\mu^2 = \lambda m_V^2$. Experimental data of pp are taken from 
	Refs.\cite{LHCb:2014acg,LHCb:2016oce,LHCb:2018rcm} while pPb are from \cite{ALICE:2014eof,ALICE:2018oyo,ALICE:2023mfc}.}
	\label{fig:rapdist_jpsi_dghp}
\end{figure}

In Fig. \ref{fig:photonproton_dghp} we present the energy dependence of the exclusive $J/\psi$ (left panel) and $\Upsilon$ (right panel) production cross - 
section in photon-proton interactions, derived using the DGHP parametrization and different values of the hard scale $\mu^2 = \lambda m_V^2$. As 
expected from Fig. \ref{fig:gluon_dghp}, the increasing of $\lambda$ implies a stepper energy dependence of the cross - section. In particular, for $\lambda 
= 1.8$, we have a very good description of the $J/\psi$ data. In addition, such a choice implies that the current data for the rapidity distributions associated 
with  exclusive $J/\psi$ photoproduction in $pp$ and $pPb$  collisions, are quite well described, as verified in Fig. \ref{fig:rapdist_jpsi_dghp}.

\begin{figure}[!h]
	\includegraphics[width=8cm]{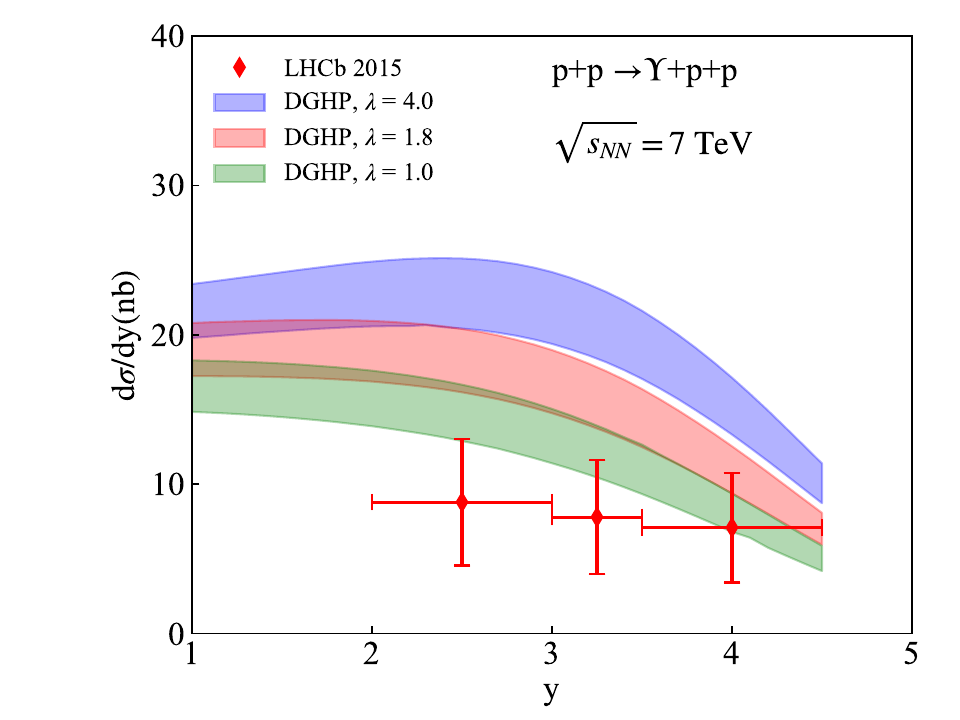}
	\includegraphics[width=8cm]{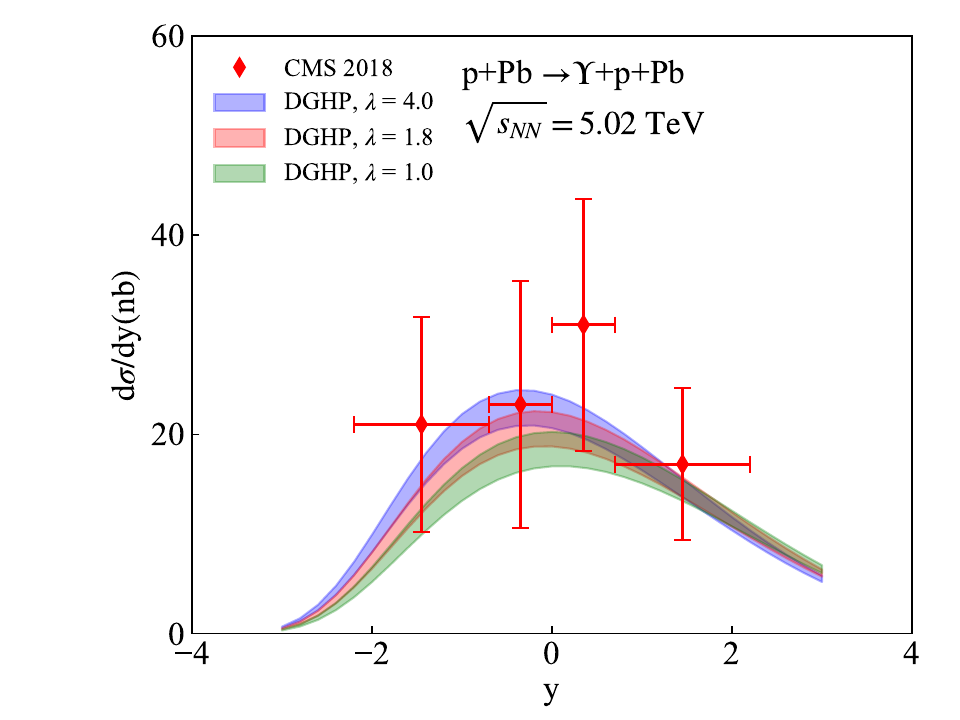}
	\caption{Rapidity distributions for the exclusive $\Upsilon$ photoproduction in $pp$ (left panel) and $pPb$ (right panel)  collisions, derived using the DGHP parametrization and different values of the hard scale $\mu^2 = \lambda m_V^2$. The experimental data are taken from 
		Ref. \cite{LHCb:2015wlx,CMS:2018bbk}.}
	\label{fig:rapdist_ups_dghp}
\end{figure}

Finally, in Fig. \ref{fig:rapdist_ups_dghp} we present our results for the rapidity distributions associated with the exclusive $\Upsilon$ photoproduction in 
$pp$ and $pPb$ collisions, derived using the DGHP parametrization and different values of the hard scale $\mu^2 = \lambda m_V^2$. In this case, we 
have that the increasing of $\lambda$ implies a worst description of $pp$ data\cite{LHCb:2015wlx}. In particular, if the value $\lambda = 1.8$ is assumed, only the data for very 
forward rapidities is described. On the other hand, the current $pPb$ data \cite{CMS:2018bbk}  are not able to discriminate the predictions derived assuming different values for the hard scale.

\section{Summary}
\label{sec:sum}

In this paper, the exclusive $J/\psi$ and $\Upsilon$ photoproduction was investigated within GPD
 approach. We have assumed the GK model and estimated the corresponding total cross - sections and rapidity distributions, considering different
  parametrizations for the unpolarized gluon distribution of the proton. In particular, three different parametrizations based on the linear DGLAP equations
  were adopted in our calculations, as well a recent parametrization that takes into account of the leading nonlinear corrections resulting from gluon
 recombination. A detailed comparison with the data was performed and the dependence of the predictions on the choice for the hard scale was estimated.
 Our results indicate that the GPD model is able to describe the current data, but a stronger conclusion about the more adequate description of the gluon
 distribution is still not possible. However, future data, in particular for the $\Upsilon$ production, could shed light on this problem and allow us to improve
 the description of the QCD dynamics at high energies.

 \section*{Acknowledgment}
 Y. P. Xie gives many thanks to Sergey Goloskokov for the communication.
  V.P.G. was partially supported by CNPq,  FAPERGS and INCT-FNA (Process No. 464898/2014-5).
This work is partially supported by the NFSC grant (Grant No. 12293061) and National Key R\&D Program of China (Grant No. 2024YFA1611000).

\end{document}